\documentclass[prl,twocolumn,showpacs]{revtex4}
\usepackage{graphicx}
\usepackage{graphics}
\usepackage{amsfonts}
\usepackage{amsmath}
\usepackage{epsfig}

\begin{document}
\catcode`\ä = \active \catcode`\ö = \active \catcode`\ü = \active
\catcode`\Ä = \active \catcode`\Ö = \active \catcode`\Ü = \active
\catcode`\ß = \active \catcode`\é = \active \catcode`\è = \active
\catcode`\ë = \active \catcode`\ô = \active \catcode`\ê = \active
\catcode`\ø = \active \catcode`\ò = \active \catcode`\í = \active
\catcode`\Ó = \active \catcode`\ú = \active \catcode`\á = \active
\catcode`\ã = \active
\defä{\"a} \defö{\"o} \defü{\"u} \defÄ{\"A} \defÖ{\"O} \defÜ{\"U} \defß{\ss} \defé{\'{e}}
\defè{\`{e}} \defë{\"{e}} \defô{\^{o}} \defê{\^{e}} \defø{\o} \defò{\`{o}} \defí{\'{i}}
\defÓ{\'{O}} \defú{\'{u}} \defá{\'{a}} \defã{\~{a}}



\newcommand{\li}{$^6$Li $ $}
\newcommand{\na}{$^{23}$Na $ $}
\newcommand{\cs}{$^{133}$Cs}
\newcommand{\kk}{$^{40}$K}
\newcommand{\rb}{$^{87}$Rb}
\newcommand{\vect}[1]{\mathbf #1}
\newcommand{\mf}{$m_F$}
\newcommand{\g}{g^{(2)}}
\newcommand{\one}{|1\rangle}
\newcommand{\two}{|2\rangle}
\newcommand{\limol}{$^6$Li$_2$}
\newcommand{\V}{V_{12}}
\newcommand{\kfa}{\frac{1}{k_F a}}

\title{Superfluid Expansion of a Rotating Fermi Gas}

\author{C.H. Schunck, M.W. Zwierlein, A. Schirotzek, and W. Ketterle}

\affiliation{Department of Physics\mbox{,} MIT-Harvard Center for
Ultracold Atoms\mbox{,}
and Research Laboratory of Electronics,\\
MIT, Cambridge, MA 02139}

\date{\today}

\begin{abstract}
We study the expansion of a rotating, superfluid Fermi gas. The
presence and absence of vortices in the rotating gas is used to
distinguish superfluid and normal parts of the expanding cloud. We
find that the superfluid pairs survive during the expansion until
the density decreases below a critical value. Our observation of
superfluid flow at this point extends the range where fermionic
superfluidity has been studied to densities of $1.2\times10^{11}$
cm$^{-3}$, about an order of magnitude lower than any previous
study.
\end{abstract}


\maketitle

\begin{figure*}
\begin{center}
\includegraphics[width=7in]{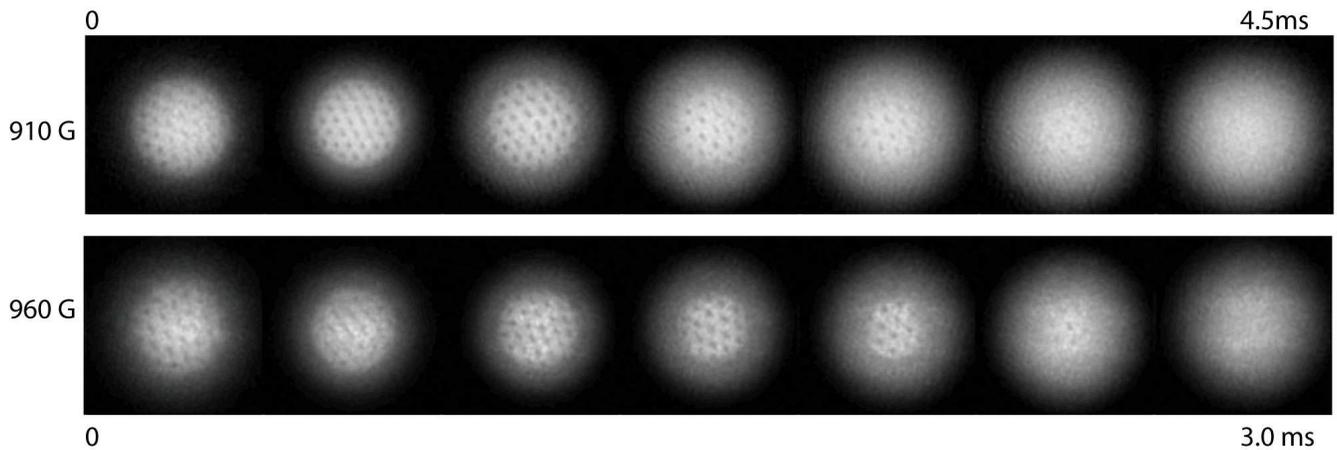}
\caption[Title]{Superfluid expansion of a strongly interacting
rotating Fermi gas. Shown are absorption images for different
expansion times on the BCS-side of the Feshbach resonance at 910 G
(0.0, 1.0, 2.0, 3.0, 3.5, 4.0, and 4.5 ms) and 960 G (0.0, 0.5, 1.0,
1.5, 2.0, 2.5, and 3 ms), before the magnetic field was ramped to
the BEC-side for further expansion. The vortices served as markers
for the superfluid parts of the cloud. Superfluidity survived the
expansion for several milliseconds and was gradually lost from the
low density edges of the cloud towards its center. Compared to 910 G
($a$ = $-7200$ $a_0$), superfluidity decayed faster at 960 G ($a$ =
$-5000$ $a_0$) due to the reduced interaction strength. The total
expansion time remained constant~\cite{pairnote3}. The field of view
of each image is 1.2 mm $\times$ 1.2 mm.} \label{fig:images}
\end{center}
\end{figure*}

Ultracold atomic gases have been used to create novel quantum
many-body systems ranging from Bose-Einstein condensates, Mott
insulators in optical lattices, to high-temperature superfluids of
strongly interacting fermions. These systems offer a high degree of
control over physical parameters including interaction strength and
density. Many important features in these gases have a spatial scale
too small to be resolved while the gas is trapped. A standard
technique to reveal this physics is to switch off the confining
potential and release the gas from the trap. A non-interacting gas
expands ballistically and the expansion reveals its momentum
distribution. The expansion dynamics of an interacting gas is
modified by the effect of collisions. This can result in classical
hydrodynamic flow and in this case the expansion serves as a (not
necessarily linear) magnifying glass for the trapped state. In
contrast to classical hydrodynamics, superfluid hydrodynamic flow
does not rely on collisions. When a weakly interacting
Bose-Einstein-Condensate (BEC) is released from an anisotropic
trapping potential, superfluid hydrodynamics leads to an inversion
of the aspect ratio, often regarded as a hallmark of Bose-Einstein
condensation~\cite{kett99var}.

The expansion dynamics of strongly interacting Fermi gases has been
the subject of a longstanding debate. For a weakly interacting
ultracold Fermi gas anisotropic expansion has been proposed as a
probe for superfluidity, analogous to the case of weakly interacting
BECs~\cite{meno02}. Anisotropic expansion has been experimentally
observed in strongly interacting Fermi
gases~\cite{ohar02science,rega03,bour03}. In this case, however, the
inversion of the aspect ratio can occur due to collisions between
the expanding atoms even if they were initially at zero
temperature~\cite{gupt04coll,jack04coll}. So far experiments have
not been able to discriminate between superfluid and collisional
hydrodynamics, and indeed one would expect both effects to
contribute: In the BCS-regime, the superfluid transition temperature
$T_C$ depends exponentially on the density. Starting at $T < T_C$,
the superfluid gas should first expand according to superfluid
hydrodynamics. As the density drops, $T$ approaches $T_C$ and
superfluidity cannot be maintained. From this point on, the gas
should expand according to collisional hydrodynamics or enter a
regime intermediate between collisional hydrodynamic and
collisionless expansion.

In this paper we study the expansion of a superfluid Fermi gas, in
the regime where pairing is purely a many-body effect. We have
observed superfluid flow even after 5 ms of expansion, when the
cloud size had increased by more than a factor of 4 and the peak
density had dropped by a factor of 17 compared to the in-trap
values.

Superfluidity in Fermi gases has previously been established through
the observation of vortex lattices~\cite{zwie05vortex,zwie06imb}. To
detect vortices in a rotating fermion pair condensate the pairs are
transferred into stable molecules by sweeping an external magnetic
field across a Feshbach resonance shortly after the gas is released
from the trap~\cite{pairnote1}. Vortices can be observed only when
the gas is still a superfluid at the moment of the magnetic field
sweep. At the final magnetic field (on the BEC side of the Feshbach
resonance) the interactions are much weaker. Therefore the vortex
core has higher contrast and is larger than near resonance. If the
gas is no longer superfluid at the time of the field ramp, we expect
the vortex core to fill in quickly and disappear. The observed
vortex cores therefore serve as markers for the regions which are
superfluid at the time of the magnetic field ramp.

Our experimental setup has been described
earlier~\cite{hadz03big_fermi}. Quantum degenerate fermionic \li was
prepared in an optical dipole trap after laser cooling and
sympathetic cooling by \na. An equal spin mixture of the two lowest
\li hyperfine states was created by a radio-frequency sweep at a
magnetic bias field of 885 G. These states, labeled $|1\rangle$ and
$|2\rangle$, exhibit a broad Feshbach resonance centered at a
magnetic field $B_0 \approx 834$ G. At magnetic fields below (above)
$B_0$, on the BEC (BCS) side, the scattering length $a$ is positive
(negative) and a nearby molecular bound state exists (does not
exist). A fermion pair condensate containing about $5 \times 10^6$
fermion pairs was obtained by evaporatively cooling the spin mixture
while ramping the magnetic field to 812 G. Note that at this
magnetic field, the bond length of the molecular state is larger
than the interatomic spacing, and the fermion pairs are bound by
many-body effects. The final radial and axial trapping frequencies
were $\omega_{r} = 2\pi \times 120$ Hz and $\omega_a = 2\pi \times
23$ Hz, respectively. To observe vortices as a probe of superfluid
flow, the gas was set in rotation: two blue-detuned laser beams were
rotated symmetrically around the cloud for 1 s at an angular
frequency of $2\pi \times 80$ Hz~\cite{zwie05vortex}. We allowed 500
ms of equilibration before the magnetic field was ramped (in 500 ms)
to several probe fields on the BCS side of the resonance. Finally,
we studied the expansion of the rotating superfluid: The gas was
released from the optical trap and expanded at the probe field for a
variable ``BCS-expansion'' time $t_{\text{BCS}}$. To transfer the
remaining fermion pairs into stable molecules the magnetic field was
then lowered in 400 $\mu$s to 680 G~\cite{pairnote2}. Here, the
cloud was given several milliseconds of ``BEC-expansion''. For
absorption imaging the magnetic field was raised to 730 G in 500
$\mu$s before the last 2 ms of time-of-flight. For most of the data
the total time-of-flight was chosen to be 11 ms~\cite{pairnote3}. An
absorption image of the gas was obtained separately at
$t_{\text{BCS}}$ to determine the peak density and the peak Fermi
momenta $k_F$ before the magnetic field sweep.

\begin{figure}
\begin{center}
\includegraphics[width=3in]{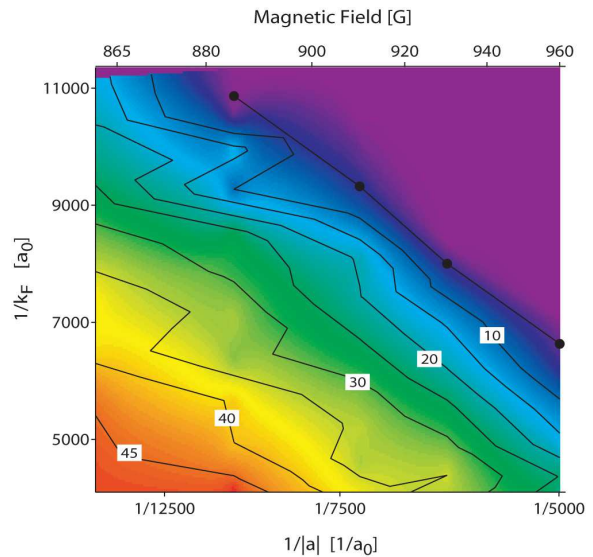}
\caption[Title]{(color online)``Phase diagram'' of an expanding,
rotating Fermi gas: At a given magnetic field the number of vortices
served as a measure for the size of the superfluid region in the
gas. The number of vortices is plotted versus $1/k_F$ and $1/|a|$.
The contour plot was created from a total of 53 data points. In this
diagram lines of constant $k_Fa$ correspond to hyperbolas. The
vortices decayed when the density (increasing $1/k_F$) or the
scattering length (increasing $1/|a|$) was reduced. For weaker
interactions, at smaller scattering lengths $|a|$, vortices were
lost already at higher densities. The four data points shown mark
the breakdown of superfluidity and are the same as those given in
Fig.~\ref{fig:kFa} (filled circles).} \label{fig:phasediagram}
\end{center}
\end{figure}

Fig.~\ref{fig:images} shows absorption images taken as outlined
above for seven different BCS-expansion times at both 910 G and 960
G. The presence of vortices proves that superfluid fermion pairs
survived in the expanding gas for several milliseconds. As the
density of the gas dropped during the BCS-expansion the vortices
were gradually lost from the low density edges of the cloud towards
its center. After 4.5 ms time-of-flight at 910 G and 3 ms at 960 G
all of the vortices had decayed. If we regard the number of vortices
as an indicator of the superfluid fraction of the gas, we can draw
the ``phase diagram'' of Fig.~\ref{fig:phasediagram}. Here the
number of vortices is shown as a function of the inverse scattering
length $1/a$ and the inverse peak Fermi momentum $1/k_F$. As $1/k_F$
increases at a given magnetic field, corresponding to the decrease
in density during time-of-flight, vortices are lost. The reduction
in the number of vortices for decreasing $|a|$ reflects the decrease
of the superfluid fraction for smaller attractive interactions at a
given temperature. In addition, the increase in the normal fraction
leads to higher damping of the remaining vortex
number~\cite{zwie05vortex}. Most importantly, however, we see that
vortices are lost earlier in time-of-flight as the interactions are
reduced.

\begin{figure}
\begin{center}
\includegraphics[width=3in]{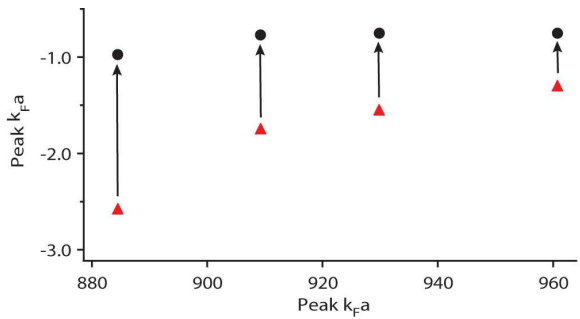}
\caption[Title]{The peak interaction strength at the end-points of
superfluid expansion. Starting at a peak $k_Fa$ in the optical trap
(red triangles), the vortices survived up to a critical peak $k_Fa$
of -0.8 +/- 0.1 (filled circles), almost independent of the magnetic
field (scattering length). The arrows mark the evolution during
expansion. The critical $k_Fa$ was obtained for each magnetic field
by taking the average of the peak $k_F$ of the last cloud that still
contained vortices at a time $t_{\text{BCS}}$ and the peak $k_F$ of
the cloud that did not contain vortices anymore 500 $\mu$s later.
The error in $k_Fa$ is about 10\% and dominated by the systematic
error in the atom number.} \label{fig:kFa}
\end{center}
\end{figure}

At all magnetic fields, we find that the peak interaction strength
at the point where all vortices were lost is about constant, $k_Fa
\sim -0.8$ (see Fig.~\ref{fig:kFa}). As shown in
Fig.~\ref{fig:images} the loss of vortices occurred gradually and
the surviving vortices were located within a circle of decreasing
radius. We assume that the critical value of $k_Fa$ for which
superfluidity was lost, was first reached at the edge of the cloud
and subsequently further inward. However, we were not able to
confirm this picture quantitatively without a model that describes
how the shape of the cloud and the bimodality develop during and
after the magnetic field sweep.

\begin{figure}
\begin{center}
\includegraphics[width=3in]{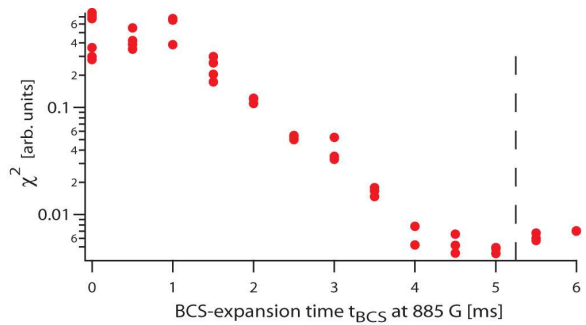}
\caption[Title]{Disappearance of bimodality. Zero temperature
Thomas-Fermi profiles were fitted to the density profiles obtained
after BCS-expansion at 885 G and subsequent BEC-expansion at 680 G.
The $\chi^2$ of the fit was monitored as a function of the
BCS-expansion time $t_{\text{BCS}}$. A high $\chi^2$ indicates a
bimodal density distribution. Vortices were still observed after 5
ms of expansion (indicated by the dashed line in the figure) while
the bimodality had already disappeared (for $\chi^2$ values smaller
than 0.01 bimodality cannot be discerned). Hence, the absence of
bimodality does not imply the absence of superfluidity.}
\label{fig:chisquare}
\end{center}
\end{figure}

\begin{figure*}
\begin{center}
\includegraphics[width=7in]{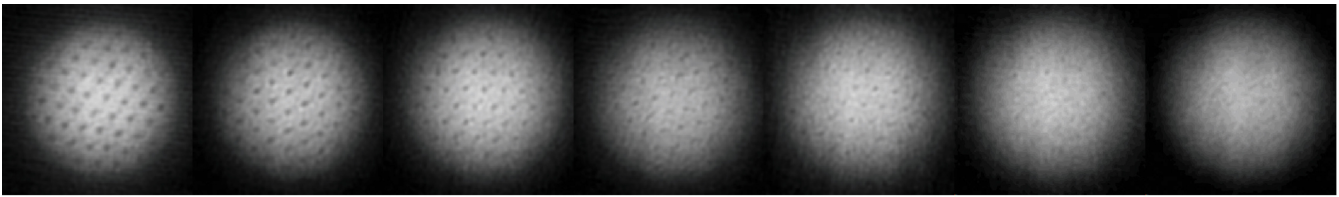}
\caption[Title]{Loss of vortex contrast on resonance at 834 G. Shown
are absorption images after a fixed total time-of-flight, but for
different expansion times on resonance (2, 2.5, 3, 3.5, 4, 5, and 6
ms) before the magnetic field was swept to the BEC-side for further
expansion. A gradual loss of the vortex contrast from about 15\%
(after 2 ms of expansion on resonance) to 3\% (after 5 ms) was
observed across the whole cloud. The field of view of each image is
1.2 mm $\times 1.2$ mm.} \label{fig:five}
\end{center}
\end{figure*}

It is remarkable that the observation of superfluidity and fermion
pair condensation for \textit{trapped} gases has also been limited
to values of $k_F|a|$ larger than
$1$~\cite{rega04,zwie04pairs,zwie05vortex}. This suggests that the
underlying reason for this limitation is the same for a trapped and
an expanding gas. One obvious scenario for the decay of the vortex
lattice during expansion is the breakdown of superfluidity when a
critical interaction strength is reached. As the density decreases,
$T_C/T_F$ drops while $T/T_F$ remains constant (since the phase
space density $n \times T^{-3/2}$ is invariant during expansion).
Therefore $T_C$ eventually becomes smaller than $T$ everywhere in
the cloud and superfluidity is lost. This scenario implies that the
superfluid state evolves adiabatically during expansion, which is
plausible: Even when the critical $k_Fa$ is reached, the pair
binding energy still changes at a slower rate,
$\dot{\Delta}/\Delta$, than the rate at which the pairs can respond
to this change, $\Delta/\hbar$~\cite{pairnote4}. Here $\Delta =
(2/e)^{7/3} E_F \exp(-\pi/2k_F|a|)$ is the pairing gap in the BCS
limit (valid for $k_F|a| \lesssim 1$)~\cite{gork61}, where the peak
Fermi energy $E_F = \hbar^2{k_F}^2/2m$ and $k_F$ are density
dependent. When superfluidity is lost, superfluid hydrodynamics is
probably replaced by an expansion intermediate between collisional
hydrodynamic and collisionless. For weakly interacting BECs, the
decay of vortex lattices at finite temperature was studied
theoretically in~\cite{krag06}, and remarkably similar structures
are found.

Another explanation for the loss of vortices is a possible failure
of the transfer of correlated fermion pairs into molecules since the
size of the fermion pairs increases with decreasing density. When
the fermion pair size becomes larger than the interparticle spacing,
molecules might be formed out of uncorrelated nearest neighbors
rather than out of correlated pairs. The magnetic field sweep then
destroys the coherent many-body wavefunction.

Vortices~\cite{zwie05vortex,zwie06imb} and bimodal density
distributions~\cite{rega04, zwie04pairs} have been used as
indicators for superfluidity and pair condensation, respectively. If
a fermion pair condensate is transferred to the BEC side before its
interaction energy has been converted into kinetic energy, it
continues to expand with the drastically reduced mean-field energy
of a molecular BEC at 680 G. This results in a clear separation of
condensate and thermal cloud after further BEC-expansion. If the
transfer of fermion pairs into molecules is delayed after releasing
the gas from the trap, the fermion pair condensate initially expands
just like the normal part of the cloud. This eventually leads to a
loss of bimodality in the density profiles after the transfer. We
can now study how the two indicators, vortices and bimodality, are
related in this experiment. For short BCS-expansion $t_{\text{BCS}}$
our data showed bimodality as well as vortices. However, the
bimodality was gradually lost and could not be discerned after a
longer BCS-expansion although vortices were still visible (see
Fig.~\ref{fig:chisquare} for details). The absence of bimodality
therefore does not indicate a breakdown of superfluidity.

So far we have studied the expansion of the gas on the BCS side of
the Feshbach resonance. On the BEC side and on resonance, $T_C$ is
proportional to $T_F$ so that $T/T_C$ is constant during expansion.
Therefore, one would not expect to observe a breakdown of
superfluidity in expansion. Fig.~\ref{fig:five} shows absorption
images that were obtained after an initial expansion of the cloud on
resonance at 834 G. In contrast to the situation on the BCS-side of
the resonance no vortices were lost. Instead, the vortex contrast
decreased uniformly across the cloud for longer expansion times.
Vortices have been detected at total densities as low as $1.2\times
10^{11}$ cm$^{-3}$ in the wings of the expanded cloud. Here the
critical temperature $T_C$ of approximately 0.2
$T_F$~\cite{bulg06,burov06} was below 20 nK ($k_B T_F$ is the local
Fermi energy). We believe that the reduction in the vortex contrast
is due to the low density of the gas after long BCS-expansion: after
the magnetic field sweep the vortex cores cannot adjust quickly
enough to the high contrast and large size they would have in
equilibrium on the BEC-side. This loss of contrast limited our study
of the breakdown of superfluidity to magnetic fields above 880 G.

In conclusion we have shown that superfluid pairs can survive during
the expansion of a strongly interacting Fermi gas. This is the first
observation of non-equilibrium superfluid flow in such systems. It
has allowed us to observe fermionic superfluidity at total densities
as low as $1.2\times 10^{11}$ cm$^{-3}$. An intriguing question for
future studies is whether fermion pairs expanding from two clouds
can coherently interfere.

We thank Gretchen Campbell for a critical reading of the manuscript.
This work was supported by the NSF, ONR, and NASA.



\begin{thebibliography}{20}
\expandafter\ifx\csname
natexlab\endcsname\relax\def\natexlab#1{#1}\fi
\expandafter\ifx\csname bibnamefont\endcsname\relax
  \def\bibnamefont#1{#1}\fi
\expandafter\ifx\csname bibfnamefont\endcsname\relax
  \def\bibfnamefont#1{#1}\fi
\expandafter\ifx\csname citenamefont\endcsname\relax
  \def\citenamefont#1{#1}\fi
\expandafter\ifx\csname url\endcsname\relax
  \def\url#1{\texttt{#1}}\fi
\expandafter\ifx\csname urlprefix\endcsname\relax\def\urlprefix{URL
}\fi \providecommand{\bibinfo}[2]{#2}
\providecommand{\eprint}[2][]{\url{#2}}


\bibitem[{\citenamefont{Ketterle et~al.}(1999)\citenamefont{Ketterle, Durfee,
  and Stamper-Kurn}}]{kett99var}
\bibinfo{author}{\bibfnamefont{W.}~\bibnamefont{Ketterle}},
  \bibinfo{author}{\bibfnamefont{D.~S.} \bibnamefont{Durfee}},
  \bibnamefont{and} \bibinfo{author}{\bibfnamefont{D.~M.}
  \bibnamefont{Stamper-Kurn}} (\bibinfo{publisher}{IOS Press},
  \bibinfo{address}{Amsterdam}, \bibinfo{year}{1999}), pp.
  \bibinfo{pages}{67--176}.

\bibitem[{\citenamefont{Menotti et~al.}(2002)\citenamefont{Menotti, Pedri, and
  Stringari}}]{meno02}
\bibinfo{author}{\bibfnamefont{C.}~\bibnamefont{Menotti}},
  \bibinfo{author}{\bibfnamefont{P.}~\bibnamefont{Pedri}}, \bibnamefont{and}
  \bibinfo{author}{\bibfnamefont{S.}~\bibnamefont{Stringari}},
  \bibinfo{journal}{Phys. Rev. Lett.} \textbf{\bibinfo{volume}{89}},
  \bibinfo{pages}{250402} (\bibinfo{year}{2002}).

\bibitem[{\citenamefont{O'Hara et~al.}(2002)\citenamefont{O'Hara, Hemmer, Gehm,
  Granade, and Thomas}}]{ohar02science}
\bibinfo{author}{\bibfnamefont{K.~M.} \bibnamefont{O'Hara}},
  \bibinfo{author}{\bibfnamefont{S.~L.} \bibnamefont{Hemmer}},
  \bibinfo{author}{\bibfnamefont{M.~E.} \bibnamefont{Gehm}},
  \bibinfo{author}{\bibfnamefont{S.~R.} \bibnamefont{Granade}},
  \bibnamefont{and} \bibinfo{author}{\bibfnamefont{J.~E.}
  \bibnamefont{Thomas}}, \bibinfo{journal}{Science}
  \textbf{\bibinfo{volume}{298}}, \bibinfo{pages}{2179} (\bibinfo{year}{2002}).

\bibitem[{\citenamefont{Regal and Jin}(2003)}]{rega03}
\bibinfo{author}{\bibfnamefont{C.~A.} \bibnamefont{Regal}} \bibnamefont{and}
  \bibinfo{author}{\bibfnamefont{D.~S.} \bibnamefont{Jin}},
  \bibinfo{journal}{Phys. Rev. Lett.} \textbf{\bibinfo{volume}{90}},
  \bibinfo{pages}{230404} (\bibinfo{year}{2003}).

\bibitem[{\citenamefont{Bourdel et~al.}(2003)\citenamefont{Bourdel, Cubizolles,
  Khaykovich, Magalhães, Kokkelmans, Shlyapnikov, and Salomon}}]{bour03}
\bibinfo{author}{\bibfnamefont{T.}~\bibnamefont{Bourdel et al.}},
  \bibinfo{journal}{Phys. Rev. Lett.} \textbf{\bibinfo{volume}{91}},
  \bibinfo{pages}{020402} (\bibinfo{year}{2003}).

\bibitem[{\citenamefont{Gupta et~al.}(2004)\citenamefont{Gupta, Hadzibabic,
  Anglin, and Ketterle}}]{gupt04coll}
\bibinfo{author}{\bibfnamefont{S.}~\bibnamefont{Gupta}},
  \bibinfo{author}{\bibfnamefont{Z.}~\bibnamefont{Hadzibabic}},
  \bibinfo{author}{\bibfnamefont{J.~R.} \bibnamefont{Anglin}},
  \bibnamefont{and} \bibinfo{author}{\bibfnamefont{W.}~\bibnamefont{Ketterle}},
  \bibinfo{journal}{Phys. Rev. Lett.} \textbf{\bibinfo{volume}{92}},
  \bibinfo{pages}{100401} (\bibinfo{year}{2004}).

\bibitem[{\citenamefont{Jackson et~al.}(2004)\citenamefont{Jackson, Pedri, and
  Stringari}}]{jack04coll}
\bibinfo{author}{\bibfnamefont{B.}~\bibnamefont{Jackson}},
  \bibinfo{author}{\bibfnamefont{P.}~\bibnamefont{Pedri}}, \bibnamefont{and}
  \bibinfo{author}{\bibfnamefont{S.}~\bibnamefont{Stringari}},
  \bibinfo{journal}{Europhys. Lett.} \textbf{\bibinfo{volume}{67}},
  \bibinfo{pages}{524} (\bibinfo{year}{2004}).

\bibitem[{\citenamefont{Zwierlein et~al.}(2005)\citenamefont{Zwierlein,
  Abo-Shaeer, Schirotzek, Schunck, and Ketterle}}]{zwie05vortex}
\bibinfo{author}{\bibfnamefont{M.~W.} \bibnamefont{Zwierlein}},
  \bibinfo{author}{\bibfnamefont{J.~R.} \bibnamefont{Abo-Shaeer}},
  \bibinfo{author}{\bibfnamefont{A.}~\bibnamefont{Schirotzek}},
  \bibinfo{author}{\bibfnamefont{C.~H.} \bibnamefont{Schunck}},
  \bibnamefont{and} \bibinfo{author}{\bibfnamefont{W.}~\bibnamefont{Ketterle}},
  \bibinfo{journal}{Nature} \textbf{\bibinfo{volume}{435}},
  \bibinfo{pages}{1047} (\bibinfo{year}{2005}).

\bibitem[{\citenamefont{Zwierlein et~al.}(2006)\citenamefont{Zwierlein,
  Schirotzek, Schunck, and Ketterle}}]{zwie06imb}
\bibinfo{author}{\bibfnamefont{M.~W.} \bibnamefont{Zwierlein}},
  \bibinfo{author}{\bibfnamefont{A.}~\bibnamefont{Schirotzek}},
  \bibinfo{author}{\bibfnamefont{C.~H.} \bibnamefont{Schunck}},
  \bibnamefont{and} \bibinfo{author}{\bibfnamefont{W.}~\bibnamefont{Ketterle}},
  \bibinfo{journal}{Science} \textbf{\bibinfo{volume}{311}},
  \bibinfo{pages}{492} (\bibinfo{year}{2006}).

\bibitem[{pai({\natexlab{b}})}]{pairnote1}
\bibinfo{note}{The sweep time is much faster than the formation time of a
  vortex lattice in the trap, which is several hundred milliseconds. Detection
  of vortices after the ramp therefore proves their presence before the ramp
  [8].}

\bibitem[{\citenamefont{Hadzibabic et~al.}(2003)\citenamefont{Hadzibabic,
  Gupta, Stan, Schunck, Zwierlein, Dieckmann, and Ketterle}}]{hadz03big_fermi}
\bibinfo{author}{\bibfnamefont{Z.}~\bibnamefont{Hadzibabic et al.}},
  \bibinfo{journal}{Phys. Rev. Lett.} \textbf{\bibinfo{volume}{91}},
  \bibinfo{pages}{160401} (\bibinfo{year}{2003}).

\bibitem[{pai({\natexlab{c}})}]{pairnote2}
\bibinfo{note}{The ramp time was 200 $\mu$s for the data taken at 834, 865 and
  885 G. We have checked that the rate of the magnetic field sweep to 680 G had
  no influence on the number of observed vortices within our measurement accuracy. }

\bibitem[{pai({\natexlab{a}})}]{pairnote3}
\bibinfo{note}{To increase the visibility of the vortices for $t_{BCS}$ = 0 and
  500 $\mu$s the total time-of-flight was increased to up to 15 ms and/or the
  power of the optical trap was increased by a factor of 4.5 during the last 2
  ms of trapping [8]}.

\bibitem[{\citenamefont{Bulgac et~al.}(2006)\citenamefont{Bulgac, Drut, and
  Magierski}}]{bulg06}
\bibinfo{author}{\bibfnamefont{A.}~\bibnamefont{Bulgac}},
  \bibinfo{author}{\bibfnamefont{J.~E.} \bibnamefont{Drut}}, \bibnamefont{and}
  \bibinfo{author}{\bibfnamefont{P.}~\bibnamefont{Magierski}},
  \bibinfo{journal}{Phys. Rev. Lett.} \textbf{\bibinfo{volume}{96}},
  \bibinfo{pages}{090404} (\bibinfo{year}{2006}).

\bibitem[{\citenamefont{Burovski et~al.}(2006)\citenamefont{Burovski,
  Prokof'ev, Svistunov, and Troyer}}]{burov06}
\bibinfo{author}{\bibfnamefont{E.}~\bibnamefont{Burovski}},
  \bibinfo{author}{\bibfnamefont{N.}~\bibnamefont{Prokof'ev}},
  \bibinfo{author}{\bibfnamefont{B.}~\bibnamefont{Svistunov}},
  \bibnamefont{and} \bibinfo{author}{\bibfnamefont{M.}~\bibnamefont{Troyer}},
  \bibinfo{journal}{Phys. Rev. Lett.} \textbf{\bibinfo{volume}{96}},
  \bibinfo{pages}{160402} (\bibinfo{year}{2006}).

\bibitem[{\citenamefont{Regal et~al.}(2004)\citenamefont{Regal, Greiner, and
  Jin}}]{rega04}
\bibinfo{author}{\bibfnamefont{C.~A.} \bibnamefont{Regal}},
  \bibinfo{author}{\bibfnamefont{M.}~\bibnamefont{Greiner}}, \bibnamefont{and}
  \bibinfo{author}{\bibfnamefont{D.~S.} \bibnamefont{Jin}},
  \bibinfo{journal}{Phys. Rev. Lett.} \textbf{\bibinfo{volume}{92}},
  \bibinfo{pages}{040403} (\bibinfo{year}{2004}).

\bibitem[{\citenamefont{Zwierlein et~al.}(2004)\citenamefont{Zwierlein, Stan,
  Schunck, Raupach, Kerman, and Ketterle}}]{zwie04pairs}
\bibinfo{author}{\bibfnamefont{M.~W.} \bibnamefont{Zwierlein et al.}},
  \bibinfo{journal}{Phys. Rev. Lett.} \textbf{\bibinfo{volume}{92}},
  \bibinfo{pages}{120403} (\bibinfo{year}{2004}).

\bibitem[{pai({\natexlab{d}})}]{pairnote4}
\bibinfo{note}{We find $\frac{\hbar\dot{\Delta}}{\Delta^2} =
  \frac{\hbar}{\Delta}\times\frac{\dot{n}}{n} \times
  \left(\frac{2}{3}+\frac{\pi}{6k_F|a|}\right)$. Since the gas expands to a
  very good approximation only radially, we assume that the density varies as
  $n(t) = n_0/(1+\omega_r^2t^2)$, and obtain
  $\left|\frac{\hbar\dot{\Delta}}{\Delta^2}\right| =
  \frac{\hbar}{\Delta}\frac{2\omega_r^2t}{1+\omega_r^2t^2}
  \left(\frac{2}{3}+\frac{\pi}{6k_F|a|}\right)$. For
  the experimental parameters when the vortices in the center of
  the cloud are lost we find that $\hbar\dot{\Delta}/{\Delta}^2 \leq 0.4$.}

\bibitem[{\citenamefont{Gor'kov and Melik-Barkhudarov}(1961)}]{gork61}
\bibinfo{author}{\bibfnamefont{L.~P.} \bibnamefont{Gor'kov}} \bibnamefont{and}
  \bibinfo{author}{\bibfnamefont{T.~K.} \bibnamefont{Melik-Barkhudarov}},
  \bibinfo{journal}{Sov. Phys. JETP} \textbf{\bibinfo{volume}{13}},
  \bibinfo{pages}{1018} (\bibinfo{year}{1961}).

\bibitem[{\citenamefont{Kragset et~al.}()\citenamefont{Kragset, Babaev, and
  Sudbø¸}}]{krag06}
\bibinfo{author}{\bibfnamefont{S.}~\bibnamefont{Kragset}},
  \bibinfo{author}{\bibfnamefont{E.}~\bibnamefont{Babaev}}, \bibnamefont{and}
  \bibinfo{author}{\bibfnamefont{A.}~\bibnamefont{Sudbø}},
  \bibinfo{note}{preprint, cond-mat/0604416}.

\end{thebibliography}

\end{document}